\documentclass[fleqn,useAMS]{mn2e}
\usepackage{graphicx}
\usepackage{url}
\usepackage{multirow}

\newcommand{\lya}{Ly$\alpha$}

\newcommand{\apjs}{ApJS}
\newcommand{\apj}{ApJ}

\newcommand{\araa}{ARA\&A}

\newcommand{\aap}{A\&A}
\newcommand{\mnras}{MNRAS}

\def\ltsima{$\; \buildrel < \over \sim \;$}
\def\simlt{\lower.5ex\hbox{\ltsima}}
\def\gtsima{$\; \buildrel > \over \sim \;$}
\def\simgt{\lower.5ex\hbox{\gtsima}}

\title[The primordial abundance of Deuterium]
{A new, precise measurement of the primordial abundance of Deuterium
\thanks{
Based on observations collected at the European Organisation for Astronomical Research 
in the Southern Hemisphere, Chile [VLT programme ID 085.A-0109(A)]. 
}}

\author[Pettini and Cooke]{Max Pettini$^{1,2}$\thanks{email: pettini@ast.cam.ac.uk} and 
Ryan Cooke$^{1,2}$ \\
$^1$Institute of Astronomy, Madingley Road, Cambridge, CB3 0HA\\
$^2$Kavli Institute for Cosmology, Madingley Road, Cambridge, CB3 0HA\\
}

\begin{document}

\date{Accepted . Received ; in original form }
\pagerange{\pageref{firstpage}--\pageref{lastpage}} 
\pubyear{2012}

\maketitle

\label{firstpage}

\begin{abstract}
The metal-poor ($Z \simeq 1/100 Z_\odot$) damped Lyman alpha 
system (DLA) at redshift $z_{\rm abs} = 3.04984$ in the $z_{\rm em} \simeq 3.030$
QSO SDSS~J1419$+$0829 has near-ideal properties for an accurate
determination of the primordial abundance of deuterium, (D/H)$_{\rm p}$.
We have analysed a high-quality spectrum of this object with software 
specifically designed to deduce the best fitting value of D/H and to assess
comprehensively the random and systematic errors affecting this
determination. We find 
(D/H)$_{\rm DLA} = (2.535 \pm 0.05) \times 10^{-5}$
which in turn implies
$\Omega_{\rm b,0} h^2 =  0.0223 \pm 0.0009$,
in very good agreement with 
$\Omega_{\rm b,0} h^2 {\rm (CMB)} =  0.0222 \pm 0.0004$
deduced from the angular power spectrum
of the cosmic microwave background. 
If the value in this DLA is indeed the true
(D/H)$_{\rm p}$ produced by Big-Bang nucleosynthesis
(BBN), there may be no need to invoke
non-standard physics nor early astration of D to
bring together $\Omega_{\rm b,0} h^2 {\rm (BBN)}$ and
$\Omega_{\rm b,0} h^2 {\rm (CMB)}$.
The scatter between most of the 
reported values of (D/H)$_{\rm p}$ in the literature
may be due largely to unaccounted systematic
errors and biases. Further progress in this area will
require a homogeneous set of data comparable to those
reported here and analysed in a self-consistent manner.
Such an endeavour, while observationally demanding,
has the potential of improving our understanding of BBN physics,
including the relevant nuclear reactions, and the subsequent processing
of $^4$He and $^7$Li through stars. 
\end{abstract}

\begin{keywords}
quasars: absorption lines -- quasars: individual: J1419+0829 -- cosmology: observations.
\end{keywords}

%%%%%%%%%%%%%%%%%%%%%%
\section{Introduction}
%%%%%%%%%%%%%%%%%%%%%%

%%%%%%%%%%%%
% FIGURE 1 %
%%%%%%%%%%%%
\begin{figure*}
  \centering
  \includegraphics[angle=0,width=17.0cm]{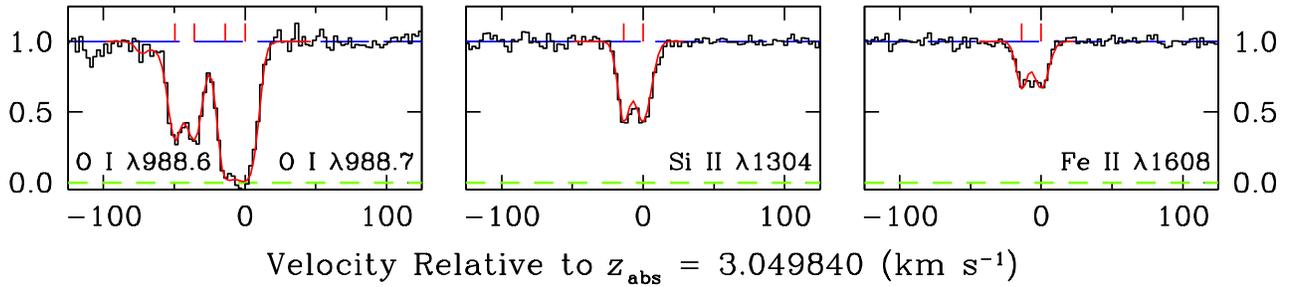}
  \caption{ 
Selected metal lines in the $z_{\rm abs} = 3.04984$ DLA in the QSO
SDSS~J1419$+$0829, reproduced from Cooke et al. (2011). In
each panel, the black histogram is the observed spectrum and the
red continuous line is the theoretical line profile fitted to the data.
Vertical tick marks above the spectrum indicate the velocities
of the two absorption components, with parameters listed in
section~\ref{sec:obs}. The $y$-axis scale is residual intensity. 
The normalized quasar continuum and zero level are 
shown by the blue long-dashed and green dashed lines, respectively.
  }
  \label{fig:metal_lines}
\end{figure*}

It is well known that the relative abundances of the light elements created
in the first few minutes of cosmic history depend on 
the product of the present-day density of baryons 
in units of the critical density, $\Omega_{\rm b,0}$,
and $h^2$, where $h$ is
the present-day value of the Hubble parameter measured in units of
100\,km~s$^{-1}$~Mpc$^{-1}$.
When analysed in conjunction with the the power spectrum
of temperature anisotropies of the cosmic background radiation (CMB),
light element abundances can improve constraints 
on cosmological parameters, primarily the spectral index of primordial
fluctuations (Pettini et al. 2008), and on the effective
number of light fermion species, $N_{\rm eff}$ (e.g. Simha \& Steigman 2008;
Nollett \& Holder 2011).

Among the elements created in Big Bang nucleosynthesis (BBN), three have
been the focus of most observational efforts: $^4$He, D and $^7$Li.
It is a lingering concern that D is the only one of the three whose
primordial abundance is consistent, within the errors, with the 
value of $\Omega_{\rm b,0} h^2$ deduced from the analysis of the
CMB. The poor sensitivity of the primordial abundance 
of $^4$He %$Y_{\rm p}$, 
to $\Omega_{\rm b,0} h^2$ requires
it to be measured with a precision which may be 
beyond what can be realistically achieved,
at least until all the sources of systematic error are fully understood
(Molaro 2008; Aver, Olive \& Skillman 2012). 
On the other hand,
a satisfactory explanation for the 
factor of $\sim 3$ discrepancy
between the abundances of $^7$Li 
in the oldest stars in our Galaxy and the value expected from BBN 
has yet to be found (Sbordone et al. 2010;  Iocco 2012; 
Nissen \& Schuster 2012 and references therein).

Even D is not entirely without its problems. 
As presciently pointed out by Adams (1976),
its primordial abundance is most effectively 
deduced from % measurements of 
the D/H ratio
in metal-poor hydrogen clouds giving rise to 
absorption lines in the spectra of high
redshift quasars (QSOs). 
This is a difficult measurement, requiring not only 
the combined power of 8--10\,m 
aperture optical telescopes and high resolution echelle spectrographs,
but also the right combination of redshift, hydrogen column 
density, and internal velocity structure of the absorbing gas.
Consequently, there are currently only about 10 QSO
absorbers where the D\,{\sc i} lines have been unambigously
detected free from interference from unrelated
absorption. 
The mean value so derived,
$\langle \log ({\rm D/H})_{\rm p} \rangle = -4.556 \pm 0.034$
(Fumagalli et al. 2011) implies 
$\Omega_{\rm b,0} h^2 {\rm (BBN)} = 0.0213 \pm 0.0012$
which agrees within the errors with 
%the value deduced from the CMB 
$\Omega_{\rm b,0} h^2 {\rm (CMB)} = 0.0222 \pm 0.0004$
(Keisler et al. 2011). 
Less satisfactory is the fact that even this prime set of
D/H measures exhibits a wider dispersion about the mean 
than expected on the basis of the quoted errors. While
the most straightforward explanation of the scatter
is that systematic errors affecting some, or all, of
the D/H measurements have been underestimated,
more radical ideas involving early destruction of
D have also been proposed (e.g. Olive et al. 2012).

Given the key role played by D in our understanding
of BBN, it is important to identify and study more QSO
absorbers with the optimum characteristics for the
determination of the primordial D abundance (D/H)$_{\rm p}$.
This is the subject of the present paper.

\subsection{Metal-poor damped Lyman alpha systems}

Among the different classes of QSO absorbers, 
the so-called `damped Lyman alpha systems' (DLAs) 
are potentially the best targets for reliable measures 
of (D/H)$_{\rm p}$. 
DLAs are characterised by high column densities
of neutral gas, with
$N$(H\,{\sc i})\,$\geq 2 \times 10^{20}$\,atoms~cm$^{-2}$
(Wolfe et al. 2005), and appear to be associated with
galaxies at early stages of chemical evolution
(Pontzen et al. 2008; P{\'e}roux et al. 2012; Krogager et al. 2012).
Typical metallicities at redshifts $z = 2$--4 are 
$Z_{\rm DLA} \simeq 1/10$--$1/100 \, Z_\odot$ (Ellison et al. 2012),
but with a tail that extends to $Z_{\rm DLA} < 1/1000 \, Z_\odot$
(Penprase et al. 2010; Cooke et al. 2011).

The number of known DLAs with metallicities less
than 1/100 of solar has increased in recent years 
thanks to large scale sky surveys (Noterdaeme et al. 2009),
and it is these systems that have proved to be 
choice candidates for measuring (D/H)$_{\rm p}$
for the following reasons:  (i) the low metallicities 
imply negligible D astration (Romano et al. 2006; Prodanovi{\'c}  et al. 2010), 
justifying the assumption that (D/H)$_{\rm DLA} = {\rm (D/H)}_{\rm p}$;
(ii) the high H\,{\sc i} column densities result in 
detectable optical depths of D\,{\sc i} absorption
in \textit{many} 
lines in the Lyman series. This both
increases the accuracy of
the measurement of $N$(D\,{\sc i}) compared with
lower column density systems, where D\,{\sc i} absorption
is detected only in the strongest 
Lyman lines (e.g. Burles \& Tytler 1998a,b),
and reduces the likelihood of contamination by unrelated
absorption; (iii) the empirical
correlation between metallicity
and absorption line width (Murphy et al. 2007;
Prochaska et al. 2008) implies that 
the kinematics of $Z \simlt 1/100 \, Z_\odot$ DLAs
are normally quiescent, with the absorption lines extending
over a narrow velocity interval (Cooke et al. 2011).
Such kinematics greatly facilitate resolving
the $-82$\,km~s$^{-1}$ isotope shift of the 
D\,{\sc i} Lyman lines from the neighbouring
H\,{\sc i} absorption.

Thus, several of the very metal-poor DLAs in the 
recent survey by Cooke et al. (2011) are promising
candidates for further observations targeting the 
high order D\,{\sc i} and H\,{\sc i} Lyman lines
redshifted into the ground-base UV spectral region
($\lambda_{\rm obs} < 4000$\,\AA).
In this paper we report the analysis of what we consider
to be the most suitable case in the sample by Cooke et al.
(2011) for determining the primordial abundance
of D with a higher precision than is normally achievable.

%%%%%%%%%%%%
% FIGURE 2 %
%%%%%%%%%%%%
\begin{figure*}
  \centering
  \includegraphics[angle=0, width=17.0cm]{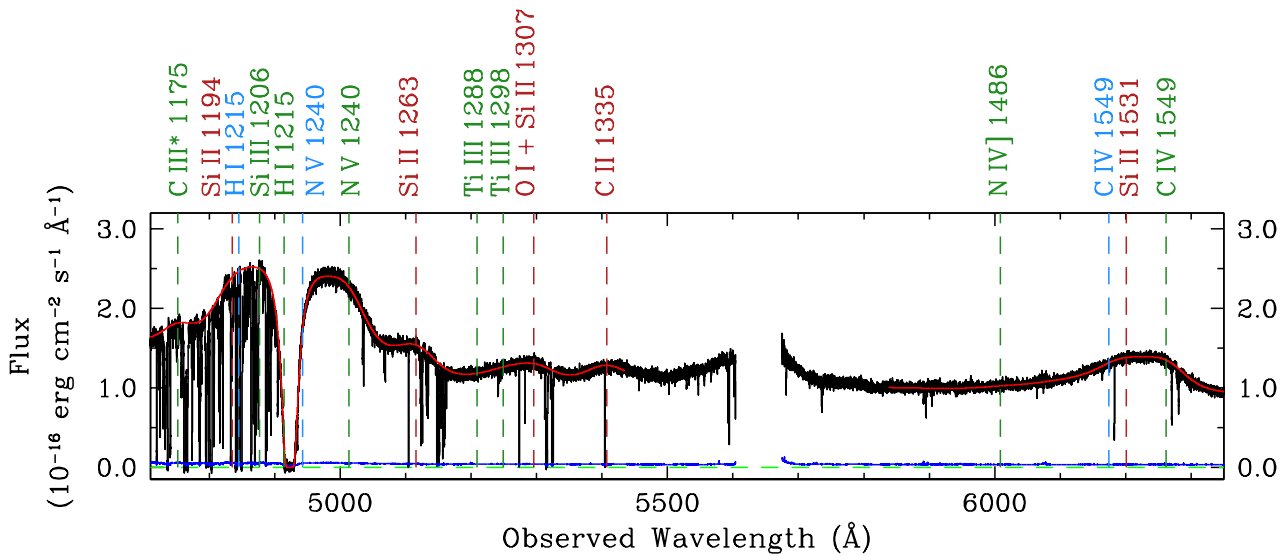}
  \caption{ 
Portion of the UVES spectrum of the QSO
SDSS~J1419$+$0829 (black), together with the 
model fit (red). 
The $1 \sigma$ error spectrum is shown in blue (near the zero level).
Vertical dash lines mark the positions
of QSO spectral features, as indicated.
Light blue labels denote emission lines at $z_{\rm em} = 2.98576$,
green labels emission lines at $z_{\rm em} = 3.04224$,
and red labels emission lines at $z_{\rm abs} = 3.04954$.
 }
  \label{fig:QSOfit1}
\end{figure*}

The paper is arranged as follows. 
In section~\ref{sec:obs} we briefly describe the observations
(which have already been reported by Cooke et al. 2011)
and summarise the most important properties of the DLA.
Sections~\ref{sec:DtoH} and \ref{sec:errors} are devoted
to the derivation of the D/H ratio and its error, respectively.
In section~\ref{sec:cosmo} we examine the cosmological
implications which follow if the value of D/H deduced
here is indeed the best measure of the primordial abundance
of deuterium. We comment on earlier measurements
of (D/H)$_{\rm p}$  in section~\ref{sec:others},
before concluding in section~\ref{sec:conc}.

\section{Observations and data reduction}
\label{sec:obs}

The spectrum of the $g = 18.9$\,mag 
QSO SDSS~J1419$+$0829 
at $z_{\rm em} \simeq 3.030$ 
shows a damped Lyman alpha
system at $z_{\rm abs} = 3.04984$; the small redshift
difference between Ly$\alpha$ emission and absorption
results in an unusually `clean' \lya\ absorption profile,
with a red wing which is essentially free from overlapping
absorption by intergalactic gas
(see Fig.~7 of Cooke et al. 2011). Consequently,
the column density of H\,{\sc i} is particularly well determined.

Details of the acquisition and reduction 
of the spectrum of this QSO can be found 
in Cooke et al. (2011); here, we summarise the 
most important points. The spectrum
was recorded with the Ultraviolet and 
Visual Echelle Spectrograph (UVES; Dekker et al. 2000) 
mounted on UT 2 of the Very Large Telescope
facility. The observations were carried out
in service mode and consisted of a series
of 2700\,s exposures; the total integration time was
29\,800\,s. 
The echelle spectra were reduced  
with the standard UVES pipeline
and then combined with the software
package \textsc{uves\_popler}\footnote{\textsc{uves\_popler} 
is available from\\
http://astronomy.swin.edu.au/$\sim$mmurphy/UVES\_popler} 
which merges individual echelle orders and maps
the data onto a vacuum heliocentric wavelength scale.
Flux calibration was achieved by reference to the 
Sloan Digital Sky Survey spectrum of the QSO.

The final reduced spectrum 
covers the wavelength range $\lambda_{\rm obs} = 3710$--6652\,\AA\
($\lambda_0 = 916$--1642\,\AA\ in the rest frame of the DLA)
with resolving power $R \simeq 40\,000$ 
(velocity resolution  FWHM\,$\simeq 7.5$\,km~s$^{-1}$)
sampled with $\sim 3$\,pixels.
The signal-to-noise ratio (S/N) of the data is relatively high 
considering the magnitude of the source and the
high spectral dispersion: ${\rm S/N} = 43$ at 5000\,\AA,
and ${\rm S/N} \geq 20$ over the entire DLA Lyman series
(these are values per pixel in the continuum).

Cooke et al. (2011) deduced 
$\log N{\rm (H\,{\textsc i})/cm}^{-2} = 20.40 \pm 0.03$.
Their analysis of 19 transitions
of N\,{\sc i}, O\,{\sc i}, Si\,{\sc ii}, and Fe\,{\sc ii}
showed that the metal lines
consist of two absorption components of approximately
equal optical depth, separated
by $\Delta v = 13.8$\,km~s$^{-1}$ at
$z_{\rm abs} = 3.049649$ and 3.049835 respectively.
The Doppler parameters 
of the two components 
are $b = 3.5$ and 6.4\,km~s$^{-1}$ respectively,
where $b = \sqrt{2} \sigma$, and $\sigma$ is
the 1-D velocity dispersion of the absorbers projected along the 
line of sight. 
The metallicity of the DLA is approximately 1/100
of solar: [O/H]\,$ = -1.92 \pm 0.05$,
[Si/H]\,$ = -2.08 \pm 0.03$, and 
[Fe/H]\,$ = - 2.33 \pm 0.04$ in the usual
notation whereby [X/H]$_{\rm DLA} = \log {\rm (X/H)}_{\rm DLA} - \log {\rm (X/H)}_\odot$.
In Fig.~\ref{fig:metal_lines} we have reproduced
examples of the metal lines in the DLA.

\section{Measurement of D/H}
\label{sec:DtoH}

For the reasons outlined above,
the spectral characteristics of this DLA are
near-ideal for an accurate determination of 
the D/H ratio and a realistic assessment of its
precision. To take advantage of this opportunity,
we have performed a detailed spectral analysis
in which we identify all of the parameters
which potentially affect the derivation of the D/H ratio
from the UVES spectrum of J1419$+$0829, and then use a
purpose-built code to solve \textit{for all of them
simultaneously} (via a $\chi^2$ statistic)
and thereby deduce the best fitting value of D/H
in the DLA. Random and systematic errors affecting
this determination were estimated using Monte Carlo techniques.
We now discuss the whole procedure in detail.

%%%%%%%%%%%%
% FIGURE 3 %
%%%%%%%%%%%%
\begin{figure}
  \centering
 {\hspace{-0.35cm} \includegraphics[angle=0, width=8.0cm]{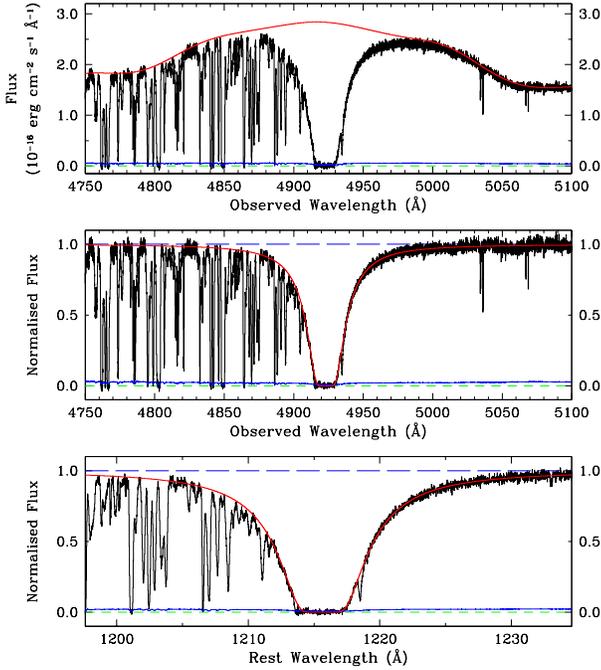}}
  \caption{ 
The \lya\ region in J1419$+$0829.
\textit{Top panel:}~Observed spectrum in black and 
best-fitting model QSO spectrum in red.
\textit{Middle panel:}~The normalized  spectrum, 
obtained by dividing the observed spectrum by the model spectrum,
is shown in black together with the 
best fitting damped \lya\ absorption profile 
(see section~\ref{sec:abs}) in red.
The neutral hydrogen column density is
$\log N{\rm (H\,\textsc{i})/cm^{-2}} = 20.391 \pm 0.008$.
\textit{Bottom panel:}~Expanded central portion of the middle panel
shown in the rest frame of the $z_{\rm abs} = 3.04984$ DLA.
In all three panels the $1 \sigma$ error spectrum is shown in blue.
 }
  \label{fig:QSOfit2}
\end{figure}

\subsection{QSO continuum}
\label{sec:QSOcont}

The absorption lines in the DLA are seen against the 
continuum and emission line radiation from the QSO;
the exact level of this flux as a function of wavelength 
affects directly the measured line optical depths.
Reconstruction of the QSO intrinsic spectrum
is particularly important for the H\,{\sc i} \lya\ 
absorption line which: (i) is blended with the QSO
\lya\ and N\,{\sc v}~$\lambda 1240$ emission lines and (ii)
affords the best constraint on $N$(H\,{\sc i}).
We approximated the QSO flux in the wavelength
regions of interest by modelling the QSO intrinsic
spectrum with a power-law continuum on which are
superimposed a number of emission lines (see Fig.~\ref{fig:QSOfit1}).
The free parameters that were varied to arrive at the optimum fit
(i.e. the fit giving the minimum value of $\chi^2$) are: 
(i) the slope and (ii) scaling constant of the power-law continuum,
and (iii) the redshifts, (iv) amplitudes and (v) widths of the emission lines.
Multiplet wavelengths for the emission lines were taken
from Morton (2003).

We found that three sets of emission line redshifts were
required to reproduce adequately the QSO emission line
spectrum: a component at $z_{\rm em} = 3.04224$
is seen in emission lines of ionised species (C\,{\sc iii},
C\,{\sc iv}, N\,{\sc iv},  N\,{\sc v}, Si\,{\sc iii}, Ti\,{\sc iii})
and in \lya; a lower redshift component at $z_{\rm em} = 2.98576$
further contributes to high ionisation lines (C\,{\sc iv} and N\,{\sc v}) 
and to \lya,
while emission lines from low ionisation species
(C\,{\sc ii}, O\,{\sc i}, and Si\,{\sc ii}) 
are centred at $z_{\rm em} = 3.04954$.
The final fit is shown in red in Fig.~\ref{fig:QSOfit1},
where the locations of the different emission line components are
colour-coded.
The region around Si\,{\sc iv}\,$\lambda 1397$ 
was not included in the fit because this emission line falls in a gap
between the CCD chips on the UVES detector.
% and no absorption lines of interest for the present work occur in that wavelength interval.
Fig.~\ref{fig:QSOfit2} shows the region near the 
\lya\ emission+absorption composite 
before and after division by the QSO flux.

Further fine adjustments were applied to the 
continuum level in the proximity of the absorption
lines to be analysed by fitting low order polynomials
to spectral intervals deemed to be free of absorption.

\subsection{Absorption lines}
\label{sec:abs}

\begin{figure*}
  \centering
  \includegraphics[angle=0, width=16.0cm]{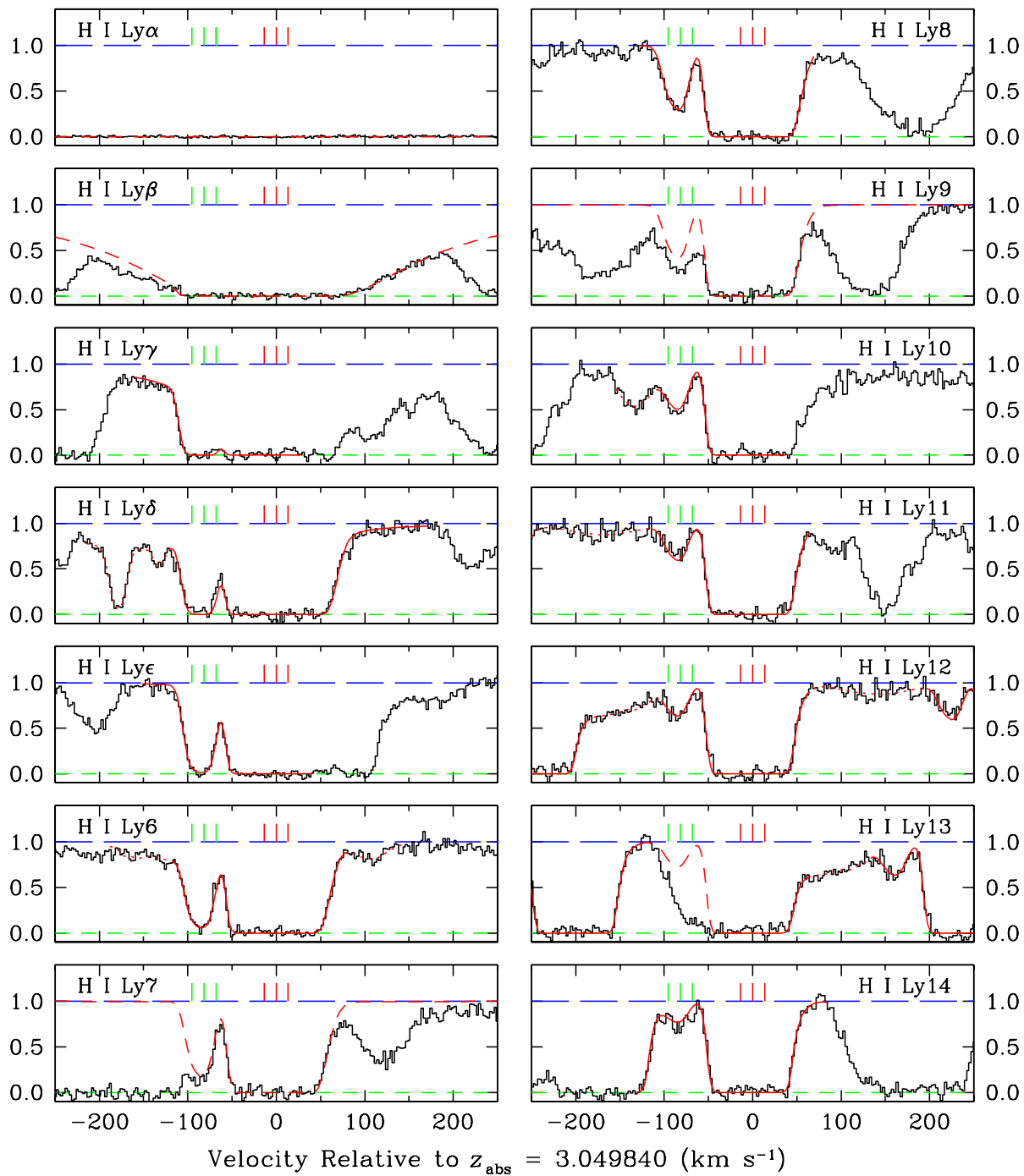}
  \caption{ 
Lyman series lines in the $z_{\rm abs} = 3.04984$ DLA.
Transitions to higher energy levels than Ly14 are too closely  
spaced in wavelength to be resolved.
The black histogram is the observed spectrum, 
while the red continuous line is the model fit to the 
absorption features (see section~\ref{sec:abs}). 
Vertical tick marks above the spectrum indicate 
the three absorption components contributing
to the H\,{\sc i} (red) and D\,{\sc i} (green)
absorption.
For completeness, we also show (dashed red line)
the model H\,{\sc i} and D\,{\sc i}
absorption superposed on blended lines
which were \emph{not} used to constrain 
the model parameters.
Dotted red lines are used for fits to features
partly blended with D\,{\sc i} and H\,{\sc i} Lyman
lines.
In all panels, the $y$-axis scale is residual intensity. 
 }
  \label{fig:Dlines}
\end{figure*}

Our UVES spectrum of J1419$+$0829
includes 19 metal absorption lines 
associated with the $z_{\rm abs} = 3.04984$ DLA; 
the analysis of these features and ensuing abundance 
determinations have been reported by Cooke et al. (2011).
Also covered by our spectrum is the entire Lyman series
of the DLA, from \lya\ to the Lyman limit. 
Beyond Ly14 (i.e. for transitions between the ground state
and levels with principal quantum number $n > 15$) 
the line wavelengths differ by $\Delta \lambda_0 < 0.5$\,\AA\
and can no longer be resolved from one another.
However, in 8 out of the 14 transitions from \lya\ to Ly14,
the D\,{\sc i} absorption can be separated from the corresponding
H\,{\sc i} and is clear of other blends
(see Fig.~\ref{fig:Dlines}).
The $f$-values of the transitions in which
D\,{\sc i} is available for analysis range from 
0.013940 (Ly$\delta$) to 0.000469 (Ly14);
this factor of $\sim 30$ spread in line strength
provides a wide baseline on the curve of growth 
for an accurate measurement of $N$(D\,{\sc i}).

%%%%%%%%%%%%
% FIGURE 4 %
%%%%%%%%%%%%

\subsubsection{Absorption components}

While only two absorption components
are evident in the metal lines analysed by Cooke et al. (2011;
see Fig.~\ref{fig:metal_lines}),
an additional, redshifted component was found to
contribute to the H\,{\sc i} absorption. 
This component, separated by $+ 12.7$\,km~s$^{-1}$
from the redshift of the DLA, has only 
$3.5 \times 10^{-4}$ the column density of the DLA,
which explains why its presence cannot be
discerned in the metal lines (recall that in this DLA
${\rm O/H} \simeq 5 \times 10^{-6}$ and other
metals are even less abundant). Even though its
contribution to the D\,{\sc i} absorption is minimal,
%we included it in the analysis for completeness.
this additional component is included in the D\,{\sc i}
lines by construction since, as explained below,
our fitting procedure
solves directly for the D\,{\sc i}/H\,{\sc i} ratio,
rather than separately for $N$(D\,{\sc i}) and 
$N$(H\,{\sc i}) in individual components. 

The absorption from each component is characterised 
by four parameters: (i) redshift $z$, (ii) column density $N$, 
(iii) temperature $T$, and (iv) large-scale velocity dispersion.
The last of these is assumed to be the same for all 
atoms and ions in the DLA, and is quantified by
a turbulent Doppler parameter $b_{\rm turb}$.
On the other hand, a given temperature
will correspond to different thermal
Doppler parameters for elements of different masses
according to the relation $b_{\rm T}^2 = 2 k T/m$
where $k$ is the Boltzmann constant.
The two contributions to the line widths add in
quadrature: 
$b_{\rm tot}^2 = b_{\rm turb}^2 + b_{\rm T}^2$.
The observation that the two components are unresolved
in the D\,{\sc i} lines while they are partially resolved
in the metal lines (compare Figs.~\ref{fig:metal_lines}
and \ref{fig:Dlines}) reflects the larger values of
$b_{\rm T}$ for the lighter D compared to the metals,
and leads to the conclusion that in this DLA thermal
broadening makes a non-negligible contribution to the
line width.

Thus, in fitting the absorption line spectrum of the DLA,
we varied simultaneously $z$, $N$, $T$, and $b_{\rm turb}$
for each of the two main  components
of the H\,{\sc i},  D\,{\sc i} and metal lines.  
(For the third component seen only in H\,{\sc i}
we varied  $z$, $N$, and $b_{\rm tot}$).
The parameters defining the QSO continuum
in the proximity of the \lya\ emission line
were also varied simultaneously to find the 
set of all of these values which minimises the 
$\chi^2$ statistic between observed and computed 
line profiles and the corresponding best-fitting value of 
D/H.\footnote{
An additional free parameter which was allowed to vary
in the $\chi^2$ minimisation procedure is the 
spectral resolution. The nominal value measured
from the widths of emission lines from the Th-Ar lamp
used for wavelength calibration is 
$R \equiv \lambda/\Delta \lambda \simeq 40\,000$.
However, since the spectrograph slit may be illuminated 
slightly differently by the lamp and the QSO, 
we preferred to treat the resolution as a free parameter,
rather than assuming that it is known perfectly.
Reassuringly, the value which was found to fit best 
the shapes of the absorption lines 
is $R = 39500 \pm 700$, as expected.
}

\begin{table*}
\centering
\begin{minipage}[c]{0.9\textwidth}
    \caption{\textsc{Parameters of best fitting absorption line model}}
    \begin{tabular}{@{}crrrrrrrrr}
    \hline
%    \hline
   \multicolumn{1}{c}{Comp.}
& \multicolumn{1}{c}{$z_{\rm abs}$} 
& \multicolumn{1}{c}{$T$} 
& \multicolumn{1}{c}{$b_{\rm turb}$} 
& \multicolumn{1}{c}{$\log N$\/(H\,{\sc i})}
& \multicolumn{1}{c}{$\log {\rm (D\,\textsc{i}/H\,\textsc{i})}$}
& \multicolumn{1}{c}{$\log N$\/(N\,{\sc i})}
& \multicolumn{1}{c}{$\log N$\/(O\,{\sc i})}
& \multicolumn{1}{c}{$\log N$\/(Si\,{\sc ii})}
& \multicolumn{1}{c}{$\log N$\/(Fe\,{\sc ii})}\\
    \multicolumn{1}{c}{}
& \multicolumn{1}{c}{}
& \multicolumn{1}{c}{(K)}
& \multicolumn{1}{c}{(km~s$^{-1}$)}
& \multicolumn{1}{c}{(cm$^{-2}$)}
& \multicolumn{1}{c}{}
& \multicolumn{1}{c}{(cm$^{-2}$)}
& \multicolumn{1}{c}{(cm$^{-2}$)}
& \multicolumn{1}{c}{(cm$^{-2}$)}
& \multicolumn{1}{c}{(cm$^{-2}$)}\\
  \hline
1  & $3.049840$            & $11\,300$               & $5.3$          & $20.231$         & $-4.601^{\rm a}$         &  $12.96$                  & $14.90$                & $13.595$                 & $13.33$ \\
    &  $\pm 0.000002$  & $\pm 200$              &$\pm 0.2$   & $\pm 0.008$   & $\pm 0.008$                &  $\pm  0.02$           & $\pm 0.01$          & $\pm 0.009$          & $\pm 0.02$ \\
2  & $3.049654$            & $10\,000$               & $2.3$          & $19.88$           & $-4.601^{\rm a}$         &  $12.99$                  & $14.81$                & $13.52$                 & $13.23$ \\
    &  $\pm 0.000001$   & $\pm 100$             &$\pm 0.2$   & $\pm 0.02$     & $\pm 0.008$                &  $\pm  0.02$           & $\pm 0.02$          & $\pm 0.02$           & $\pm 0.03$ \\
3  & $3.0500$               & \ldots$^{\rm b}$   & $26$           & $16.9$             & $-4.601^{\rm a}$         &  \ldots$^{\rm b}$  & \ldots$^{\rm b}$ & \ldots$^{\rm b}$   & \ldots$^{\rm b}$ \\
    &  $\pm 0.0001$      &                               &$\pm 2$      & $\pm 0.3$       & $\pm 0.008$  \\
  \hline
    \end{tabular}
    \smallskip
$^{\rm a}${Fixed to be the same in all three components.}\\
$^{\rm b}${The third component is undetected in the metal and D\,{\sc i} lines.}\\
    \label{tab:best_fit}
\end{minipage}
\end{table*}

It is worth stressing here that this procedure is
somewhat different from that adopted in some of the 
previous derivations of the deuterium abundance.
Specifically, in most previous studies of D/H in DLAs, 
the column densities of H\,{\sc i}
and D\,{\sc i} were determined separately,
the former being largely derived from the damping wings 
of the \lya\ line and the latter from the available
D\,{\sc i} lines in the Lyman series.
The ratio of these two column
densities then gives D/H.
In the present case, by fitting simultaneously
all of the H\,{\sc i}, D\,{\sc i}
and metal lines, as well as the QSO flux from which
the absorption takes place,
we used \textit{all} of the information
available in the DLA absorption spectrum
to deduce directly the value of D/H
(assumed to be the same in all three absorption
components) which best fits the data.

In order to fit the data as explained, we developed 
custom-made software based on the same principles
as \textsc{vpfit}, the software package most commonly used in the
analysis of interstellar absorption lines (see 
http://www.ast.cam.ac.uk/${\sim}$rfc/vpfit.html).
Our \textsc{absorption line fitting (alfit)} code,
written in the \textsc{python} programming language,
uses a modified version
of the \textsc{mpfit} package (Markwardt 2009).\footnote{ 
\textsc{mpfit} was originally written for the \textsc{interactive data
language (IDL)} environment and has recently been converted
into \textsc{python} by Mark Rivers and Sergey Koposov.}
\textsc{mpfit} employs a Levenberg-Marquardt
technique to derive the set of model parameters that minimises 
the $\chi^2$ statistic.
We tested \textsc{alfit} extensively using fake data
with known model parameters, and against \textsc{vpfit}
in simpler cases than that described here, to ensure
that its output is fully compatible with that of 
previous analyses which used \textsc{vpfit}.

The parameters of the best fitting model are collected
in Table~\ref{tab:best_fit}, and the corresponding 
line profiles are shown superposed on 
the observed spectral features in Fig.~\ref{fig:Dlines}.
It is interesting to note that
the two components that
make up the DLA 
(components 1 and 2 in Table~\ref{tab:best_fit}) 
have temperatures $T \simeq 11\,000$ and 10\,000\,K
respectively. A comparably high temperature
has been reported recently by Carswell et al. (2012)
for a similarly metal-poor DLA, while in other cases
somewhat lower temperatures have been inferred
(Pettini et al. 2008; Cooke et al. 2012).
Temperatures of a few thousand degrees for
metal-poor DLAs are consistent with the 
lack of 21\,cm (Srianand et al. 2012; Ellison et al. 2012) 
and molecular hydrogen
(Petitjean et al. 2000) absorption in such systems.
We are not able to derive a value of the temperature for 
component 3 because it is detected only in H\,{\sc i} 
and our method relies on the comparison of the widths of
absorption lines from elements of differing masses.

The column densities listed in Table~\ref{tab:best_fit}
differ by only $\sim 0.02$--0.03\,dex from the values 
reported by Cooke et al. (2011) whose simpler analysis 
did not differentiate between turbulent and thermal 
broadening of the absorption lines. For the total 
neutral hydrogen column density in the DLA we obtain
$\log N{\rm (H\,\textsc{i})/cm}^{-2} = 20.391 \pm 0.008$.
The best-fitting value of D/H is 
$\log {\rm (D\,\textsc{i}/H\,\textsc{i})} = -4.601 \pm 0.008$.

\section{Error determination}
\label{sec:errors}

%%%%%%%%%%%%
% FIGURE 5 %
%%%%%%%%%%%%
\begin{figure}
  \centering
  \hspace{-0.25cm}\includegraphics[angle=0, width=7.5cm]{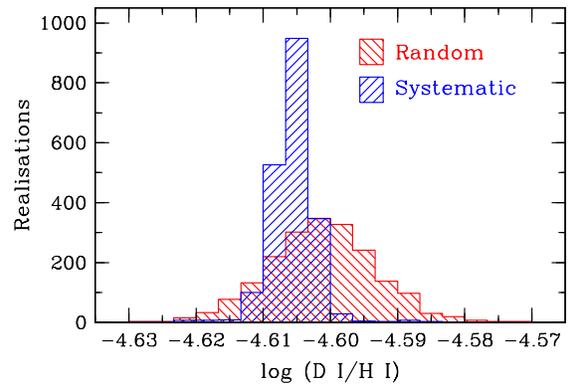}
  \caption{Histograms showing the distribution of values
  of $\log {\rm D\,\textsc{i}/H\,\textsc{i}}$ 
  in 2000 Monte Carlo random realisations of the spectrum of 
  J1419$+$0829, as described in section~\ref{sec:errors}.
  }
  \label{fig:error_hist}
\end{figure}

An important aspect of the present analysis is an in-depth
assessment of the errors, both random and systematic, 
which apply to the value of D/H deduced above. 
The first thing to point out is that the errors listed
in Table~\ref{tab:best_fit} are purely random,
reflecting the effect that the $1 \sigma$ error spectrum
has on the derivation of the model parameters;
these errors are just the diagonal terms from
the covariance matrix.

%%%%%%%%%%%%
% FIGURE 6 %
%%%%%%%%%%%%
\begin{figure*}
  \centering
  \includegraphics[angle=0, width=15.0cm]{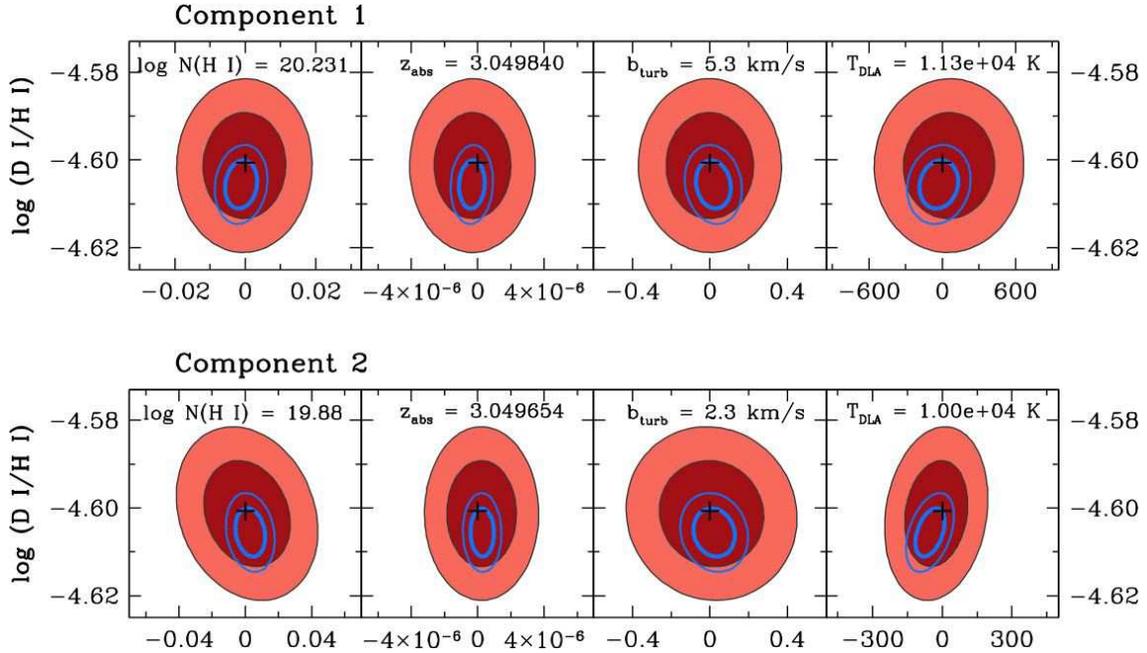}
  \caption{Elliptical contour plots illustrating the covariance between
  model parameters and the
  D\,{\sc i}/H\,{\sc i} ratio. 
 The black cross near the centre of each panel shows the best-fitting
 value of each parameter, as indicated (see also Table~\ref{tab:best_fit}),
 while the $x$-axis gives the deviation from this value.
 The red-shaded contours refer to the 1 and $2 \sigma$ ranges
 for the random errors; the corresponding ranges for the systematic
 errors are shown in blue.  
 }
  \label{fig:contours}
\end{figure*}

We tested the validity of this estimate of
the random error applicable to the best fitting value
of D/H
using a suite of simulations, as follows.
The parameters defining the best fitting model
were used to generate a fake UVES spectrum
which was then perturbed by a random realisation
of the error spectrum.
To ensure that our choice of starting parameters
in the line fitting procedure 
does not bias the result, for each realisation
we randomly drew a new set of starting parameters
from the covariance array of the best-fitting
model that describes the real data.
The resulting spectrum was
then analysed in an identical manner to the real
spectrum (as described in section~\ref{sec:DtoH}),
to derive a new set of best fitting parameters,
including the D/H value. This procedure was
repeated 2000 times and the distribution of resulting values
examined to infer the $1 \sigma$ random
error on D/H 
(i.e. the range about the mean encompassing 
68\% of the 2000 returned values).  
We found that  the error on 
$\log {\rm (D\,\textsc{i}/H\,\textsc{i})}$ so derived
is $1 \sigma_{\rm ran} = \pm 0.008$ (see Fig.~\ref{fig:error_hist}), 
in good agreement with the value 
obtained from the diagonal terms of the covariance matrix.

The two major sources of systematic error are
the continuum placement 
% (or flux calibration for the case
% of the DLA's \lya\ absorption line) 
and the zero level; 
both affect the measured optical depths of the
absorption lines.
To assess their impact on the derived value
of D/H, we used the above 2000 Monte Carlo realisations and
applied continuum and zero level adjustments that reflect 
the operations performed on the real data. In this way, 
we can also identify any systematic bias that may be
introduced as a 
result of our continuum and zero level placements.

Examination of the 
black cores of saturated H\,{\sc i} lines
of the DLA's Lyman series
indicated that the zero level:
(i) is independent of wavelength, 
and (ii) has an uncertainty of $\pm 0.0013$ ($1 \sigma$)
of the continuum flux.
For the 2000 realisations described above, we
perturbed the zero level
by a constant amount,
drawn at random from a Gaussian distribution of
values about zero with a dispersion of 0.0013.
% The spectrum was then renormalised following the 
% exact same steps as for the real data, and
% the best fitting model parameters redetermined.
The uncertainty resulting from the continuum placement
was estimated by repeating the same fine continuum adjustments
described in section~\ref{sec:QSOcont} on each of the
2000 randomly generated spectra and 
redetermining the best fitting model parameters as before.

For each realisation, we take the difference between the 
best-fitting value of D/H, respectively with and without the 
aforementioned corrections, % to the continuum and zero level placement, 
as an indication of the systematic uncertainty
affecting the derivation of the D/H ratio.
The distribution of such differences in the 2000 
trials then gives the estimate of the 
systematic error and bias in our measure
of $\log {\rm (D\,\textsc{i}/H\,\textsc{i})}$. 

As can be seen  from Fig.~\ref{fig:error_hist},
this second set of tests indicates that: (i) 
the magnitude of the systematic error 
is small: $1 \sigma_{\rm sys} = \pm 0.003$;
and (ii) there appears to be a bias of 
$- 0.005$ in our estimate of 
$\log {\rm (D\,\textsc{i}/H\,\textsc{i})}$
introduced by a (small) error in the continuum
placement. 
A further advantage of our approach is that it allows us 
to examine in detail the sensitivity 
of the D/H ratio to
each of the model parameters entering in its derivation.
In Fig.~\ref{fig:contours} we have reproduced
some examples; the error contours are `well-behaved'
and we see only mild correlations between
D\,\textsc{i}/H\,\textsc{i} and some of the model
parameters.

The result of all of the above tests is as follows:
\begin{equation}
\log {\rm (D\,\textsc{i}/H\,\textsc{i})} = -4.596 \pm 0.008 \pm 0.003 \, 
\label{eq:D1H1a}
\end{equation}
(random and systematic $1 \sigma$ errors respectively),
which includes the correction of $+0.005$\,dex
for the bias in the continuum and zero level placement.

Combining the two errors in quadrature, leads to:
\begin{equation}
\log {\rm (D\,\textsc{i}/H\,\textsc{i})} = -4.596 \pm 0.009\, ,
\label{eq:D1H1b}
\end{equation}
or:
\begin{equation}
10^5\, {\rm (D\,\textsc{i}/H\,\textsc{i})} = 2.535 \pm 0.05\, .
\label{eq:D1H1c}
\end{equation}

\section{Cosmological implications}
\label{sec:cosmo}

Our determination of D/H in J1419$+$0829 is more precise than
those reported up to now in other QSO absorbers for the following
reasons: (i) The red wing of the damped \lya\ line suffers very little
contamination by gas unrelated to the DLA, easing the 
determination of $N$(H\,{\sc i}); 
(ii) eight D\,{\sc i} Lyman lines of widely differing
$f$-values are accessible; (iii) the kinematic structure
of the gas is simple, with only 2--3 components
contributing to the absorption lines; and 
(iv) the spectrum analysed is of moderately high S/N.
% and (v) we have paid particular attention to the determination
% of the errors with a suite of Monte Carlo simulations.
We therefore consider it worthwhile to examine the 
cosmological implications of the new measurement reported
here before discussing the full sample of available
D/H measures at high redshift. 

In the following, we take:
\begin{equation}
{\rm (D\,\textsc{i}/H\,\textsc{i})}_{\rm DLA} = {\rm (D/H)_{\rm DLA}} = {\rm (D/H)_{\rm p}} \,.
\label{eq:D1H1d}
\end{equation}
The assumptions underlying these equalities are that:
(i) the fractional ionizations of H and D are the same,
(ii) D is not depleted relative to H, and (iii) 
the destruction of D through astration prior
to the time when we observe the DLA has been
negligible. 
Concerning the first assumption, we are not aware of
a physical process that would under-
or over-ionise one isotope relative to the other.
Dust depletion of D in the local interstellar 
medium has been proposed to explain the 
surprising range of D/H values found along different sightlines in
our Galaxy (Linsky et al. 2006), but is unlikely
to be important in metal- and dust-poor DLAs 
where even highly refractory elements are present
in near-solar relative proportions 
(Akerman et al. 2005; Vladilo et al. 2006; Ellison et al. 2007).
Given the low metallicity of the $z_{\rm abs} = 3.04984$ DLA,
where N, O, Si, and Fe have abundances less than
$\sim 1/100$ of solar (Cooke et al.  2011), the third assumption 
is supported by chemical evolution
models which entertain 
little reduction of the D abundance from its
primordial value when such a 
small fraction of the gas 
has evidently been cycled through stars
(Romano et al. 2006).

Recently, Steigman (private communication)
has updated the relations between (D/H)$_{\rm p}$
and $\Omega_{\rm b,0} h^2$ given by Simha \& Steigman (2008)
and Steigman (2007), as follows:
\begin{equation}
10^5\, {\rm (D/H)} = 2.60 (1  \pm 0.06) 
\left [ \frac{6}{\eta_{10} - 6 (S -1)}\right ]^{1.6}\, ,
\label{eq:bar1}
\end{equation}
where $\eta_{10}$ is the post-$e^{\pm}$ annihilation
ratio of the numbers of baryons and photons in units of $10^{-10}$:
\begin{equation}
\eta_{10} \equiv 10^{10} n_{\rm b}/n_{\gamma} = 273.9 \, \Omega_{\rm b,0} \, h^2 \, ,
\label{eq:eta10}
\end{equation}
and $S$ is the non-standard expansion rate factor, related
to the number of additional (equivalent) neutrinos
$\Delta {\rm N}_\nu \equiv {\rm N}_\nu -3$ by:
\begin{equation}
S^2 \equiv \left ( \frac{H^\prime}{H}\right )^2 = 1 + \frac{7 \Delta {\rm N}_\nu}{43}
\label{eq:S}
\end{equation}
The 6\% error in the conversion factor in eq.~(\ref{eq:bar1})
reflects the  uncertainties in the nuclear reaction rates,
particularly the $d(p,\gamma)^3{\rm He}$ cross-section
(e.g. Nollett \& Holder 2011), used in BBN codes (Steigman, 
private communication).

For standard BBN with $S = 1$, 
eqs.~(\ref{eq:D1H1c}), (\ref{eq:bar1}), and (\ref{eq:eta10})
yield:
\begin{equation}
100 \Omega_{\rm b,0}  h^2 {\rm (BBN)} = 2.23 \pm 0.03 \pm 0.08
\label{eq:OmegaBBN1}
\end{equation}
where the error terms reflect the uncertainties in,
respectively, (D/H)$_{\rm p}$ (eq.~\ref{eq:D1H1c})
and BBN calculations (eq.~\ref{eq:bar1}).
Combining the two error terms in quadrature, we have:
\begin{equation}
100 \Omega_{\rm b,0}  h^2 {\rm (BBN)} = 2.23 \pm 0.09 \, .
\label{eq:OmegaBBN2}
\end{equation}

%%%%%%%%%%%%
% FIGURE 7 %
%%%%%%%%%%%%
\begin{figure}
%\vspace*{-3.75cm}
  \centering
  {\hspace*{-0.00002cm}\includegraphics[angle=0,width=80mm]{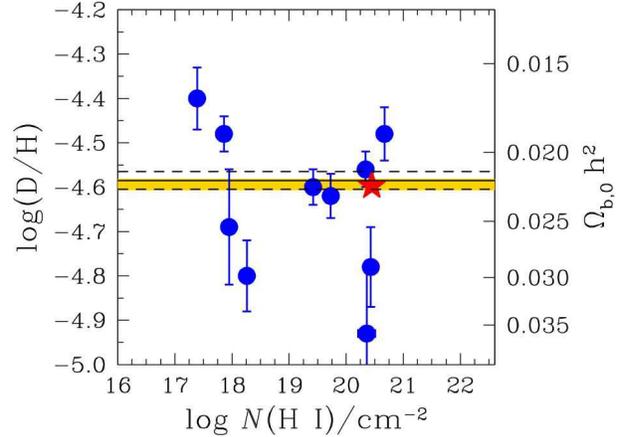}}
%\vspace{-4.05cm}
  \caption{
   Measures of the deuterium abundance in high redshift QSO absorbers.
   Only cases were the deuterium absorption is clearly resolved from
   nearby spectral features are shown (see text). 
   % Blue circles
   % denote systems observed from the ground with 8-10\,m telescopes
   % and echelle spectrographs, while the red triangle refers to lower
   % resolution observations made with the \textit{Hubble Space Telescope}.
   % Absorption systems discussed in the text are labelled with the name
   % of the background QSO. 
   The red star refers to the new measurement reported here,
   with errors smaller than the symbol size.
   The horizontal lines are drawn at the
   weighted mean value of $\log {\rm (D/H)}$
   and its error, as determined with the bootstrap method.
   The yellow shaded area shows the range in 
   $\Omega_{\rm b,0} h^2 ({\rm CMB})$ from Keisler et al. (2011).
   }
   \label{fig:DtoH}
\end{figure}

Recently, Keisler et al. (2011) combined their 
measurement of the CMB angular power spectrum 
from the South Pole Telescope (SPT) with the power spectra
from the seven-year Wilkinson Microwave Anisotropy Probe (WMAP) 
data release to better constrain cosmological parameters.
From this analysis, it was concluded:
\begin{equation}
100 \Omega_{\rm b,0}  h^2 {\rm (CMB)} = 2.22 \pm 0.042 \, 
\label{eq:OmegaCMB}
\end{equation}
(see Table~3 of Keisler et al. 2011).

The agreement between eqs.~(\ref{eq:OmegaBBN2}) and (\ref{eq:OmegaCMB})
is very encouraging.
If the value of (D/H)$_{\rm p}$ we have deduced here is indeed
the correct one, there may be no need to appeal to non-standard physics
to reconcile BBN and CMB estimates of $\Omega_{\rm b,0}$
(see, for example, the discussion of this point by Hamann et al. 2011).
Specifically, if we adopt the value 
$100 \Omega_{\rm b,0} h^2 {\rm (CMB)} = 2.22 \pm 0.042$
from Keisler et al. (2011) 
and then use eqs.~(\ref{eq:bar1}), (\ref{eq:eta10}), and (\ref{eq:S})
to solve for $\Delta {\rm N}_\nu$ using the determination of (D/H)$_{\rm p}$
reported here (eq.~\ref{eq:D1H1c}), we find:
\begin{equation}
{\rm N}_\nu = 3.0 \pm 0.5
\label{eq:N_nu}
\end{equation}
combining in quadrature the errors on (D/H)$_{\rm p}$, BBN reaction rates
and $\Omega_{\rm b,0} h^2 {\rm (CMB)}$, with the BBN rates
making the largest contribution to the total uncertainty.
This value is in good agreement with ${\rm N}_\nu = 2.984 \pm 0.008$
deduced from the width of Z-boson decays in electron-positron colliders
(ALEPH Collaboration et al. 2006).

\begin{table*}
\centering
\begin{minipage}[c]{0.7\textwidth}
    \caption{\textsc{Prime Sample of D/H measurements in QSO Absorption Line Systems}}
    \begin{tabular}{@{}lrrrrrr}
    \hline
%    \hline
   \multicolumn{1}{c}{QSO}
& \multicolumn{1}{c}{$z_{\rm em}$} 
& \multicolumn{1}{c}{$z_{\rm abs}$} 
& \multicolumn{1}{c}{$\log N$\/(H\,{\sc i})}
& \multicolumn{1}{c}{[O/H]$^{\rm a}$}
& \multicolumn{1}{c}{$\log {\rm (D/H)}$}
& \multicolumn{1}{c}{Ref.$^{\rm b}$}\\
    \multicolumn{1}{c}{}
& \multicolumn{1}{c}{}
& \multicolumn{1}{c}{}
& \multicolumn{1}{c}{(cm$^{-2}$)}
& \multicolumn{1}{c}{}
& \multicolumn{1}{c}{}
& \multicolumn{1}{c}{}\\    
  \hline
HS\,0105+1619                        & 2.640     &  2.53600   & $19.42 \pm 0.01$   & $-1.73$                   &  $-4.60 \pm 0.04$    &  1 \\
Q0913+072                            & 2.785     &  2.61843   & $20.34 \pm 0.04$   & $-2.40$                   &  $-4.56 \pm 0.04$    &  2, 3 \\
Q1009+299                            & 2.640     &  2.50357   & $17.39 \pm 0.06$   & $< -0.70^{\rm c}$          &  $-4.40 \pm 0.07$    &  4 \\
SDSS~J1134$+$5742                    & 3.522     &  3.41088   & $17.95 \pm 0.05$   & $< -1.9^{\rm d}$           &  $-4.69 \pm 0.13$    &  5 \\
Q1243+307                            & 2.558     &  2.52566   & $19.73 \pm 0.04$   & $-2.79$                   &  $-4.62 \pm 0.05$    &  6 \\
SDSS~J1337$+$3152                    & 3.174     &  3.16768   & $20.41 \pm 0.15$   & $-2.68$                   &  $-4.93 \pm 0.15$    &  7 \\
SDSS~J1419$+$0829                    & 3.030     &  3.04984   & $20.391 \pm 0.008$   & $-1.92$                   &  $-4.596 \pm 0.009$    &  8, 9 \\
SDSS~J1558$-$0031                    & 2.823     &  2.70262   & $20.67 \pm 0.05$   & $-1.50$                   &  $-4.48 \pm 0.06$    &  10  \\
Q1937$-$101                          & 3.787     &  3.25601   & $18.26 \pm 0.02$   & $ -2.0^{\rm e}$                 &  $-4.80 \pm 0.08$    &  11 \\
Q1937$-$101                          & 3.787     &  3.57220   & $17.86 \pm 0.02$   & $ < -0.9$                 &  $-4.48 \pm 0.04$    &  12 \\
Q2206$-$199                          & 2.559     &  2.07624   & $20.43 \pm 0.04$   & $-2.07$                   &  $-4.78 \pm 0.09$    &  2, 13 \\
   \hline
    \end{tabular}
    \smallskip

$^{\rm a}${Relative to the solar value $\log ({\rm O/H})_{\odot} + 12 = 8.69$ (Asplund et al. 2009).}\\
$^{\rm b}${References -- (1)~O'Meara et al. (2001), 
(2)~Pettini et al. (2008a),
(3)~Pettini et al. (2008b),
(4)~Burles \& Tytler (1998b),
(5)~Fumagalli et al. (2011),
(6)~Kirkman et al. (2003),
(7)~Srianand et al. (2010),
(8)~This work,
(9)~Cooke et al. (2011),
(10)~O'Meara et al. (2006),
(11)~Crighton et al. (2004),
(12)~Burles \& Tytler (1998a),
(13)~Pettini \& Bowen (2001).
}\\
$^{\rm c}${This is a very conservative upper limit on the metallicity. Burles \& Tytler (1998b)
estimate [Si/H]\,$\simeq -2.5$ and [C/H]\,$\simeq -2.9$ from photoionisation modelling.\\
$^{\rm d}${This is a conservative upper limit on the metallicity. Fumagalli et al. (2011)
estimate [Si/H]\,$\simeq -4.2$ from photoionisation modelling.}\\
$^{\rm e}${This value refers to [Si/H] as the O abundance was not measured by Crighton et al. (2004).}
}\\
    \label{tab:DtoH}
\end{minipage}
\end{table*}

\section{Comparison with other measurements of D/H}
\label{sec:others}

%In reality, things are not so clear-cut of course. 
%Even setting aside the problems associated with the
%primordial abundances of $^4$He and $^7$Li discussed
%earlier,
In the previous section, we used our new,
precise measure of 
D/H to compare the present day baryon density derived 
from BBN calculations and CMB measurements. 
However, a full comparison between 
$\Omega_{\rm b,0}{\rm (BBN)}$ 
and $\Omega_{\rm b,0}{\rm (CMB)}$ should 
include all reliable measurements of the primordial
abundances of the light elements created in BBN.
It is beyond the scope of this paper to comment on
the difficulties in deducing the primordial abundances
of $^4$He and $^7$Li from, respectively, the emission
line spectra of metal-poor H\,{\sc ii} regions and
the absorption spectra of some of the oldest stars 
in the Galaxy. We refer the interested reader to the 
review by Steigman (2007) for a detailed 
discussion.

Here, we focus specifically on the derivation of
(D/H)$_{\rm p}$.
Since the compilation assembled by Pettini et al. (2008),
two new measurements of (D/H)$_{\rm p}$ have been
reported, by Srianand et al. (2010) and 
Fumagalli et al. (2011). 
In the former case, the D\,{\sc i} lines are only partially
resolved from nearby blends, so that its inclusion 
in what we consider to be
the most reliable set of measurements is questionable
but, for completeness, we have included this DLA in
the sample. 
Together with the new detection reported here,
the full data set now consists of 11 measurements
of (D/H)$_{\rm p}$ whose main characteristics 
are collected in Table~\ref{tab:DtoH}.

As can be seen from Fig.~\ref{fig:DtoH}, there is
a troublesome dispersion between the 11 measurements,
well in excess of the quoted errors. 
The problem is that even this set is highly heterogeneous.
The number of D\,{\sc i} Lyman lines covered 
varies between the 11 QSO absorbers,
with only Ly$\alpha$ and Ly$\beta$ 
available in systems of lower $N$(H\,{\sc i}).
The degree of line blending is significantly worse
in some cases than in others. 
Some spectra are of inferior S/N ratio than others,
such as that of the $z_{\rm abs} = 2.07624$
DLA in Q2206$-$199 which, because of the lower
redshift of this absorber, necessitated
observations at ultraviolet wavelengths with
the \textit{Hubble Space Telescope}
(Pettini \& Bowen 2001). 
Perhaps most importantly, 
the efforts devoted to evaluating 
realistic errors vary considerably between
the different analyses collected in Table~\ref{tab:DtoH},
and we suspect that in at least some cases the errors quoted
may well be underestimates, and the values reported
may suffer from biases that
are unaccounted for.

It is our contention that the only way forward 
with this problem is to assemble a uniform sample
of metal-poor DLAs, observed at comparable S/N,
and analysed consistently following procedures
similar to those 
employed here. Unfortunately, this ambitious goal
is still some way off in the future.
Given the reservations expressed above, it is unclear
to the present authors whether there is much to be learnt 
by averaging the measurements in Table~\ref{tab:DtoH}.
Nevertheless, for completeness, we have estimated the
mean $\langle {\rm (D/H)}_{\rm p} \rangle$ using
the bootstrap method. That is, we calculated
the weighted mean of the 11 measurements in 
Table~\ref{tab:DtoH} adopting the errors quoted in the original
reports, and repeated the calculation 10000 times
by random sampling (with substitution) of the 
11 values of (D/H)$_{\rm p}$. The mean and standard
deviation of the results of 10000 such trials are:
\begin{equation}
\langle \log {\rm (D/H)}_{\rm p} \rangle = -4.585 \pm 0.02 \, .
\label{eq:mean}
\end{equation}

For comparison, Fumagalli et al. (2011) calculated 
$\langle \log ({\rm D/H})_{\rm p} \rangle = -4.556 \pm 0.034$,
while Pettini et al. (2008) reported
$\langle \log ({\rm D/H})_{\rm p} \rangle = -4.55 \pm 0.03$;
both estimates are from subsets of the data in Table~\ref{tab:DtoH}.
Thus, the new case highlighted here has reduced slightly both the
mean of the sample and its error, but the main conclusion is that the
value of $\langle \log ({\rm D/H})_{\rm p} \rangle$
has remained relatively stable over the last few years.

\section{Summary and Conclusions}
\label{sec:conc}

In summary, we have identified a metal-poor DLA from the
recent survey of such systems by Cooke et al. (2011) with 
near-ideal properties for an accurate determination of the
primordial abundance of deuterium. To capitalise on this rare 
opportunity, we have developed
a spectral analysis specifically targeted at finding the
most likely value of D/H by varying all of the 
spectral parameters
which have a bearing on this measurement.
We have also paid particular attention to the
assessment of the errors and biases, both random
and systematic, which affect our determination of D/H.

Our principal result is that the value of 
(D/H)$_{\rm p}$ so derived,
(D/H)$_{\rm p} = (2.535 \pm 0.05) \times 10^{-5}$,
implies $\Omega_{\rm b,0} h^2 =  0.0223 \pm 0.0009$
which is in excellent agreement with 
$\Omega_{\rm b,0} h^2 {\rm (CMB)} =  0.0222 \pm 0.0004$
deduced from the joint analysis of the CMB
angular power spectrum recorded with the 
WMAP and SPT experiments.
On the basis of the new measurement reported here,
there is no need to invoke non-standard physics,
nor appeal to early astration of deuterium, 
to reconcile CMB and D/H determinations 
of the cosmic density of baryons.

When considered in the context of previous
reports of D/H values in QSO absorbers,
the new case presented in this paper 
highlights the heterogeneity of the existing
data set. The weighted mean of what
are often considered to be the best 11
available measurements is
$\langle {\rm (D/H)}_{\rm p} \rangle = (2.6 \pm 0.1) \times 10^{-5}$,
but it is hard to assess the significance
of this  mean value when it is likely
that systematic errors and biases may have 
been overlooked in at least some earlier analyses.

Looking ahead, we can expect a further refinement in the
determination of $\Omega_{\rm b,0} h^2 {\rm (CMB)}$
from the Planck mission (Planck Collaboration et 
al. 2011). 
However, improvements in the value of (D/H)$_{\rm p}$
made possible by the identification and careful analysis
of suitable metal-poor DLAs remain a priority,
given their potential for clarifying our understanding
of primordial nucleosynthesis (including the relevant
nuclear reaction rates) and of the subsequent
processing of $^4$He and $^7$Li. As an added bonus,
more detections of D\,{\sc i} absorption in metal-poor
DLAs will increase the limited statistics 
on the kinetic temperature of these structures
and help us isolate the dominant   
physical processes that regulate it.

\section*{Acknowledgements}

We are grateful to the ESO 
time assignment committee
for their continuing support
of this demanding observational programme,
and to the staff astronomers at Paranal 
who conducted the observations.
It is a pleasure to acknowledge helpful 
and stimulating conversations
with  Bob Carswell, Dawn Erb, Jan Hamann, 
Paul Hewett, Mike Irwin, Lloyd Knox,
Michael Murphy, Subir Sarkar, and Gary Steigman.
We are grateful to Gary Steigman
for communicating his revised fitting formulae
in advance of publication, and to the referee, 
John Webb, for valuable comments
which improved the paper.

\label{lastpage}

\end{document}